\documentclass[12pt]{article}
\usepackage{fpcp03}

\begin{document}


\Title{Analytical methods in heavy quark physics and the case of $\tau_{1/2}(w)$}
\bigskip


\label{Le YaouancStart}

\author{A. Le Yaouanc\index{Le Yaouanc, A.} }

\address{Laboratoire de Physique Th\'eorique - UMR CNRS 8627,\\
 Universit\'e Paris XI - B\^atiment 210, F-91405 Orsay Cedex, France}

\makeauthor\abstracts{\bf  Analytical methods in heavy quark physics are shortly reviewed,
with emphasis on the problems of dynamical calculations. Then, attention is attracted to the various difficulties raised by a tentative experimental determination of $\tau_{1/2}$. }

\section{Introduction}
\label{sec1:leyaouanc}
\hspace*{\parindent} Stimulated
particularly by the perspective of elucidating {\it CP violation}, the study of heavy quark 
physics has produced in a
rather short time remarkable and unexpectedly strong results, due to a very active and close cooperation of a
community of theoreticians, and of experimentalists. Although, at some places, recourse to heavy {\it numerical} methods of {\it lattice QCD} is necessary, what is nevertheless also remarkable is the richness of {\it analytical} methods and  results, and how far we can advance by the use of rather simple means. We must be necessarily highly selective in this very short expos\'e \footnote{This is a slightly extended version of the talk delivered on the occasion of the conference FPCP2003 held at old Ecole polytechnique, Montagne Sainte-Genevi\`eve, june 2003. In addition, the sections 3.2 (quark models) and 3.3 (radial excitations) prepared for the talk could not actually be exposed.}.
In particular, we choose to concentrate mainly on the problems of {\it
heavy to heavy hadron transitions}, especially  heavy, $b \to c$, currents,
and we select the following methods: 1) symmetry and inclusive sum rules approaches 2) QCD sum rules 3) Quark models. Then, as a striking illustration, we will end by a discussion of the unexpectedly close connection which exists between the $\lambda_1$
or $\mu_{\pi}^2$ parameter and the $B \to D^{**}\pi$ experiment, through the $\tau_{1/2}$ parameter which characterises the transition $B \to D_{0^+}^{**}$. Theoretical and experimental indications on this parameter are confronted \footnote{In this corrected version (v2), we correct some numbers concerning the estimate of $tau_{1/2}$ from the $B \to D^{**}\pi$ decays (class III). We also add short information in footnotes on new experimental results(in class I) and calculations, and some quotations of authors.}.\\

\noindent \underline{\bf Partial list of authors in the field covered by the
talk} It would not make sense, and it would not be possible, in this very short report, to give fair references to the people having contributed to the subject. So we choose the unusual procedure of just giving a list of some of
the workers in the field, with due apology for the missing people : Isgur, Wise, Falk, Grinstein, Boyd, Ligeti, Leibovich, Neubert, Uraltsev, Bigi,
Ball, Braun, Bagan, Shifman, Blok, Mannel, Melikhov, Braun, Khodjamirian, Ruckl,
Neubert, Sachrajda, Buras, Buchalla, Colangelo, de Fazio, Paver,
Vainshtein, Golowich, Donoghue, Burdman, Huang, Dai,
Stech, Soares, Goity, Roberts, Eichten, Hill, Bardeen, Martinelli, Rosner, Quigg, Lebed, Luke, Ali, Gambino, Pham, Nardulli, Fleming, Stewart,
 Jaus, Wyler, Pirjol, Lellouch, Faustov, Cheng, and last, our own little community (with often D. Melikhov) :$\left \{\begin{array}{l} \hbox{Oliver, Raynal, Le Yaouanc, P\`ene}\\ 
\hbox{Becirevic,
Charles, Mor\'enas}\end{array}\right .$ 

\vskip 0.5cm

\noindent \underline{\bf Definition of quantities frequently under discussion }

- $\rho^2$: slope of $\xi (w) = - \displaystyle{{d{\xi}(w) \over
dw}}$

- $\tau_{1/2}$ resp. $\tau_{3/2}$~: form factors for current transition
from $0^-$ to $L=1$, $j=1/2$ or $3/2$ states in the $m_Q \to \infty$ limit. They control the deacys $\left ( \begin{array}{l}  B \to D^{**}\ell \nu 
\\ B
\to D^{**}\pi \end{array} \right .$

- $\overline{\Lambda} \sim m_B - m_b$ (to be simple)

- $\mu_{\pi}^2$ or $-\lambda_1 \sim \displaystyle{{1 \over 2m_B}} <B|-D^2|B>$,
and $\mu_G^2 \sim {3 \over 4} (m_{B^*}^2 - m_B^2)$ control $1/m_b^2$ corrections to $\Gamma_{B \to X_c \ell \nu}$ 

- $\rho_D^3$: controls the $1/m_b^3$ corrections to $\Gamma_{B \to
X_c\ell \nu}$\par \vskip 1 truecm

\section{A short survey}
\label{sec2:leyaouanc}

\subsection{Exact results} \label{subsec2.1:leyaouanc}

\subsubsection{Exact results at infinite mass limit ($m_Q \to \infty$)}
\label{subsubsec2.1.1:leyaouanc}

Concern both pure QCD (spectra, strong decays, ...) and hadron {\it
electroweak} properties (form factors, ...)

- {\bf Symmetries} Ex. $m_D = m_{D^*}$~; universality of form factors
$B^{(*)} \to D^{(*)}$ and the function $\xi (w)$

- {\bf Small velocity (SV) sum rules} , leading to {\it lower bounds}. Recent ones~ex.: 
$\rho^2 \geq 3/4$, for curvature $\sigma^2 = 2c \geq {4 \rho^2+3 \rho^4 \over
5}$, (exist also for all higher derivatives). \par

\noindent NB There exists also {\it finite} mass {\it dispersive bounds} on form 
factor through {\it t channel} SR.\\

- Further properties with inclusion of {\it light} mesons

\ 1) Union of $\chi$PT (chiral pert. th.) and heavy quark limit for {\it
soft} pions

\underline{Ex. VMD} (exact at $1/m_Q$ included) in $B \to \pi \ell \nu$
at $q^2_{max}$.

$\begin{array}{l} \hbox{2) $B \to \pi \ell \nu$ with}\\ \quad 
\hbox{{\it Hard} mesons}\\ \quad \hbox{$E_{\pi} \to \infty$ (LEET)} \end{array} \left\{ 
\begin{array}{l} \hbox{new symmetries Ex. $f_0 \sim E_{\pi}f_+$} \\
\hbox{asymptotic behaviour Ex. $f_+(0) \sim E_{\pi}^{-2}$}\\
\hbox{approx. factorisation of $B \to \pi \pi$
("BBNS")}\end{array}\right.$\\

- Finally, in exceptional cases, one can give {\bf absolute} predictions

\noindent For $b \to c$, $\xi (w=1) = 1 $ ; $\Gamma_{B \to X_c\ell \nu} = \Gamma _{b \to c\ell
\nu}$ perturbative (quarks + gluons) (NB The latter requires knowledge of $m_{b,c}$ and $\alpha_s$, but these are basic parameters of QCD) 

\subsubsection{Finite mass corrections ; $1/m_Q$ expansions}
\label{subsubsec2.1.2:leyaouanc} Now, to obtain physical results, one must
nevertheless know what happens at finite masses $m_b$, $m_c$. Among
simplest results

$\hbox{Ex. :}  \quad h_{A_1}(w=1) = 1 + {\cal O} \left ( {1
\over m_Q^2}\right ) + \hbox{rad. corr.}$ 

$\hbox{Ex. :} \quad \Gamma_B \to X_c \ell \nu = \Gamma_{b\to c
\ell \nu} \left ( 1 + {\cal O} \left ( {1 \over m_b^2}\right ) \right
)$

\noindent In both cases, the ``power'' corrections have coefficient
controlled by heavy meson-heavy meson matrix elements of simple
operators : at order ${1 \over m_b^2}$, one has two of them, $\mu_{\pi}^2$ (or $- \lambda_1$), $\mu_G^2$
(or $3\lambda_2$).

\subsubsection{Effective theories} \label{subsubsec2.1.3:leyaouanc} 

One can
exhibit these heavy quark limit properties in a {\it Lagrangian}
formalism, as "effective" theories, for which QCD takes very {\it simple} forms.

$$ \begin{array}{lll} \hbox{{\bf HQ}ET} &m_Q \to \infty
&\begin{array}{l}{\cal L} \sim i \overline{Q}_v v^{\mu} {\cal D}_{\mu}
Q_{v} \quad \hbox{($Q$ or $h$)}\\ \qquad \qquad \quad \hbox{$\rightarrow$ 
spin
symmetry} \end{array} \\ \hbox{{\bf LE}ET} &\begin{array}{l} m_Q \to
\infty \\ E_{\pi} \to \infty \end{array} &\widetilde{\cal L} \sim i
\overline{q}_n \not v (n^{\mu} D_{\mu})q_n \end{array}$$

$Q$ : heavy quark~; $q$ : light quark, $v$ velocity of $Q$, $n$ unit vector along  the light
meson momentum. ET is for effective theory, LEET for large energy effective theory.

\noindent HQET allows to perform {\it dynamical} calculations or {\it
perturbative} calculations in field theory in the heavy quark limit, and also to calculate the $
1/m_Q$ corrections.

\noindent [NB However, one can also work in full QCD, with {\it finite
physical masses}. This is not always more difficult. Yet heavy quark
expansion, if converging sufficiently quickly, has the advantage that it allows to concentrate on
a limited set of parameters~: $\rho^2$, $\mu_{\pi}^2$, etc .. ((almost)
independent of quark masses)].

\subsubsection{Wilson operator product expansion (OPE)}
\label{subsubsec2.1.4:leyaouanc} Let us finally recall that the OPE is central in heavy quark
theory, since it lies behind the $1/m_Q$ expansion -- for instance for
effective theories. More generally, it allows to develop Green functions in
inverse powers of a large parameter, which may be also a large {\it
momentum} (as in QCDSR at $m_Q = \infty$). The coefficients of the
expansion are matrix elements of operators between hadrons or over
vacuum, multiplied by {\it perturbative} factors.

\subsection{Dynamical calculations of hadronic matrix elements}
\label{subsec2.2:leyaouanc}

{\bf Need for dynamical calculations}. Obviously, theory of heavy
quark limit does not allow to avoid dynamical calculations save in exceptional
cases~: $B \to D^{(*)}$ at $w = 1$. Inclusive $\Gamma (B \to X_c\ell
\nu)$ at least requires basic QCD parameters, $m_b$, $m_c$ ($\overline{\Lambda}$), as well as $\alpha_s$. Matrix elements 
of currents, and related parameters like $\rho^2$,
$\xi (w)$ remain unknown. $1/m_Q$ expansion requires new unknown matrix elements like $\mu_{\pi}^2$ or $\rho_D^3$ ... to be calculated (or determined
{\it empirically} as recently by CLEO and DELPHI). Let us concentrate on matrix elements calculations.
 
\noindent {\bf Main available methods of dynamical calculation}.
- The most fundamental dynamical method (entirely based on QCD Lagrangian) is {\it lattice} QCD method. It relies on very heavy {\it numerical} calculations -- and still rather approximate. Certainly, in principle, these approximations are only approximations in the numerical method used to estimate the path integrals,
which means that they can be reduced systematically by improvement of numerical technics. Yet, drawbacks indeed exist, and they are twofold : 1) Calculations
may appear too heavy for practical reasons, ranging from financial ones to lack
of working forces willing to do the job, or too long time to await truely good
results. 2) One may personally suffer too much from the lack of physical intuition partially inherent 
to such methods. 

$$\begin{array}{l} \hbox{two mainly {\it analytical} methods} \\ \hbox{to
calculate matrix elements}\end{array} \left \{ \begin{array}{l}
\hbox{-"QCD sum rules" (QCDSR)} \\ \hbox{not to be confused with simple ``sum rules''}\\ \hbox{-quark models} \end{array}
\right . $$

\noindent These latter methods are appealing because 1) they require much less calculational efforts, and 2) especially for quark models, they provide easily
physical insight. However, the price to pay is that both involve approximations to QCD which
cannot be reduced {\it systematically}, neither can {\it the errors be
safely estimated}. These types of approximations are quite different in nature from the ones present in lattice QCD. There are additional hypotheses like to suppose the velocity to be not too large, or to suppose the radial excitations to be weakly coupled ....We can say : yes, it is so, but we must add : perhaps it is not so. And moreover, we will not know the answer in general, except by comparing with the correct result or with the experiment. Then, we have possibly to change our hypotheses, and therefore our predictions too. In short, these methods are not very predictive. In conclusion, we can say, imitating some well known statement :``Il n'y a pas de voie royale pour QCD". All the methods should rather be
considered as complementary. 

1) {\bf QCDSR}. The first step is the OPE expansion of Green functions, with extensive use of {\it perturbative} QCD plus a limited {\it non perturbative} ingredient through {\it vacuum
condensates}. This is the well (QCD-) founded side. The other, more questionable,
side is to extract the {\it ground state} properties from the
calculated Green function. This is not possible by a {\it systematic}
algorithm~; rather,it is  a {\it matter of art}, with a phenomenological
model, and with various parameters to fix~: continuum threshold,
fiducial range of the Borel parameter $M^2$, stability criteria ... The resulting accuracy
is then {\it moderate} and {\it variable} according to the problem
under study. 

2) {\bf Quark models}. Useful in view of the uncertainties remaining in more
``fundamental'' methods, lattice QCD as well as QCDSR. They have only a {\it qualitative} relation to QCD through the QCD inspired potential~:

$$V(r) = \lambda r - {4 \over 3} \ {\alpha (r) \over r}$$

- The advantage of QM is that it is founded on a direct intuition of {\it
bound state} physics, unlike QCDSR. Important since, after all, we
deal really with bound states.

- However, bound state physics is something very {\it complex} beyond non relativistic (NR) approximation. This complexity is reflected in the large {\it variety} of quark
models. There is no ``standard'' quark model. One then gets quite often a very
{\it large range} of predictions. The thing to do is then to understand
what is {\it behind} each model. It requires {\it time}. Anyway, the
accuracy in QM too will remain quite {\it variable} for possibly quite
a long time.

\subsection{Models as testing bench}
\label{subsec2.3:leyaouanc}
Before entering into more details about 
dynamical calculations, one must finally 
tell a word about another important application of {\it models}, that is using them as a {\it testing
bench} for QCD. Indeed,models may also help to analyse certain features of QCD, or problems of QCD
methods, in much {\it simpler}, although perhaps unphysical situations, where one can perform much more {\it explicit} calculations.  \par \vskip 2 truemm

Two important applications $\begin{array}{l} \hbox{- check of {\it
duality} in inclusive decays}\\ \hbox{- checks of QCDSR approach.}
\end{array}$ \par \vskip 2 truemm

Two types of models are useful : 1) `t Hooft model (QCD$_2$, $N_c \to \infty$),
which has the advantage of deriving from a covariant field theory 2) quark models are certainly theoretically cruder, but they have the advantage of working in three-dimensional space. 
The NR models satisfy {\it exact}
duality, and allows completely explicit calculations in the harmonic oscillator case. On the other hand, the relativistic {\it Bakamjian-Thomas models} are useful to check {\it leading order}
SV sum rules of QCD (see below, subsection "Quark models").

\section{Specific problems of (analytical)dynamical methods
}
\label{sec3:leyaouanc}
As we have said, a general problem of QCDSR and quark models is that,
to a different extent, they are not \underline{{\it definite}
approximations} to QCD. However, in the present context of {\it heavy}
quarks, they have more {\it specific} problems, useful to recall in view
of the discrepancies observed with {\it experiment} (See below
subsection "Consequences to be drawn from experiment").

\subsection{QCDSR method}
\label{subsec3.1:leyaouanc}
Here, the main problem seems the standard treatment of {\it radial
excitations} (usually represented through the ``{\it continuum}'' {\it model}) on the {\it hadronic}, or
phenomenological side, or ``L.H.S.''. The situation is critical in the so called ``three-point
QCDSR'', especially in HQET.

Let us indeed concentrate on the problem of {\it ground state} to {\it
ground state} matrix elements (M.E.) of an operator which is ${\cal
O} = j^{\mu}$ or $j^{\mu5}$ for $\rho^2$ and $\tau_{1/2}$~, and 
$\vec{D}^2$ in the case of $\mu_{\pi}^2$, ... (ground state is meant for lowest state in a
channel of definite $J^P$). If one looks for $<H'|O|H>$, one considers the
{\it Green function}~: $<0|T(j^{H'}, {\cal O}, j^H)|0>$. $j^{H,H'}$ are ``{\it interpolating}'' currents connecting the vacuum
to $H$ and $H'$ with factors $f_{H,H'}$, whence a contribution $\sim
f_{H}f_{H'} <H'|{\cal O}|H>$. Knowing $f_{H,H'}$ from other QCDSR (two-point), and calculating the
above Green function by OPE, one would thus manage to obtain

$$<H'|{\cal O}|H> \mathrel{\mathop {\sim}^?}  {<0|T(j^{H'}, {\cal O},
j^H)|0> \over f_H f_{H'}}$$

\noindent {\it It is not so simple however}. Indeed, $j^{H,H'}$ also connects the vacuum to {\it all}
the {\it radial} excitations $H^{(n)}$, $H'^{(n')}$. Then one {\it must
get rid} of these radial excitations. The standard way to do it is twofold~:

1) One approximates the radial states by a {\it continuum}, itself
calculated by OPE. This introduces a first basic parameter, the {\it
continuum threshold} $s_c$.

2) This approximation introduces an error. One tries to {\it reduce} it
by a {\it Borel} transformation, affecting the states of mass $s,s'$ by factors $ exp (- {s \over M^2}),exp (- {s'
\over M'^2)}$ which reduce relatively the contribution of radial excitations if $M^2$,
$M'^2$ sufficiently {\it small}. However $M^2$ must also not be {\it too}
small. Otherwise one has to calculate too much power corrections in the
OPE calculation. Whence a {\it "fiducial"} {\it window} for
$M^2$~; $M'^2$~: $M_{min}^2 < M^2,M'^2 < M_{max}^2$. \\

Now, both steps 1) and 2) {\it may fail} somewhat, precluding good
accuracy (sometimes, very large error):

- particulary in three-point sum rules in HQET, (not only) it {\it may}
happen {\bf first} that $M^2$ cannot be made sufficiently low to reduce
enough the excitation contribution ({\it rapidly rising} spectral
functions). This seems to be the case for $\mu_{\pi}^2$.

- {\bf second}, in addition, in three-point sum rules, it may happen
that the presentation of excitations by the OPE continuum is defective,
particularly because one may have strong radial contributions with {\it sign
opposite} to the ground state. This seems to be the case for the $\widehat{g} \sim g_{D^*D\pi}$
calculation. Including a negative radial excitation restores stability
in $M^2$ and makes $\widehat{g}$ {\it twice} larger, restoring
agreement with experiment. According to a NR harmonic oscillator calculation, the problem could appear in $\rho^2$
QCDSR. \par \vskip 2 truemm

  {\it Possible} symptoms of problems :
$\begin{array}{l}\hbox{$\alpha$) a lack of stability in
$M^2$ : this is present for $\mu_{\pi}^2$ } \\ \hbox{$\beta$) a too critical dependence on 
$s_c$}\end{array}$ But these symptoms may be absent, though the result be disputable.

\subsection{Quark models} \label{subsec3.2:leyaouanc} Since there is a
great {\it variety} of models, we have to make a {\it selection}.
Nevertheless, we will try to underline the {\it physical} problems
leading to the various possible choices.

1) {\bf Non relativistic (NR) model}

It is the basic one. It is a fully {\it consistent} ({\it Hamiltonian} model) and easy to use. Weaknesses~: the
ones of NR mechanics in a rather {\it relativistic} world. To be
specific 
 :
$\alpha$) NR {\it kinematics} may seem a very crude approximation for $b \to c$. Not
absurd however at {\it small recoil}~; but this is misleading if one
calculates a {\it derivative}, e.g. $\rho^2$, which is
sensitive to relativistic boost effects.

$\beta$) Still more worrying, {\it internal velocities} are not small
in {\it heavy-light} systems~: $\Delta$ (excitation energy with respect
to ground state) $\sim m \sim 0.35$~GeV. Therefore $v/c$ expansion may
be totally misleading unlike in the $\Upsilon$ system. As examples :

\quad --$\rho^2$ is predicted too small $\rho^2 \sim 0.5$

\quad --$1/m_Q^2$ corrections to $h_{A_1}(1)$ (overlap effect in
the NR approach) are {\it much} too small ; one misses the main contribution.

2) {\bf Relativistic models \`a la Bakamjian-Thomas}

This class of models includes the well known {\it light-front} quark models. They represent an important progress. They present indeed several qualities, starting from a much improved, relativistic,
treatment of the {\it center-of-mass motion}, in a full {\it
Hamiltonian} formalism. They are well adapted to $b \to c$ form factor
calculations. 

More precisely, Bakamjian-Thomas models possess the following good properties~: 

$\left \{ \begin{array}{l} \hbox{- realistic $\rho^2 \sim 1$~; corr. to
$h_{A_1}(1)$ sensible (related to $<\vec{p}^{\, 2}>$)} \\ \hbox{- {\it
covariance} in the $m_Q \to \infty$ limit, as well as with $E_{\pi} \to
\infty$ in $B \to \pi$ (LEET limit)}\\ \hbox{- symmetry relations of
HQET and LEET} \\ \hbox{- relativistic form of Bjorken and Uraltsev
sum rules as well as higher}\\ \hbox{\ \  derivative sum rules (curvature
...)}\end{array} \right . $

\noindent However they do not satisfy sum rules for higher moments, like
Voloshin sum rule.

Another defect, which may become critical~: approximation of {\it free}
quark Dirac spinors, essential to these models, may become too rough
for the {\it light} quark in $Q\overline{q}$. Ex : corrections to
$\widehat{g}(g_{D^*D\pi})$ are too large~: $\widehat{g}$ too {\it
low}.

3) {\bf Dirac equation}

This is another, complementary, answer to defects of NR model. The light quark is considered to move in the field of a
{\it static} source (quark $b$ or $c$). It offers a much improved treatment of
Dirac spinors, taking into account interaction. This allows to restore
agreement for $\widehat{g} \sim 0.6$. In fact, it is convenient for $\left \{
\begin{array}{l} \hbox{spectrum}\\
\hbox{elementary quantum $\pi$,
$\gamma$, emission from the {\it light} quark }\end{array} \right .$

\noindent However, it is {\it not} adapted to $b \to c$ (no treatment of center-of-mass motion). There exists extensions of Dirac equation to finite $m_Q$ and to {\it
moving} hadrons. However, no attempt has been made to calculate heavy current {\it form factors}.

\subsection{Decays of radial excitations} \label{subsec3.3:leyaouanc}

Finally, let us quote a problem for all these methods of dynamical calculation. Namely,
we lack a satisfactory treatment of {\it pionic} decays of {\it radial
excitations} $D'$, $B'$ ..., like $D'^{(*)} \to D^{(*)}\pi$. Indeed :

- Standard QCDSR are precisely fitted to predict only the properties
of the {\it lowest state} in each channel. 

- {\it Quark models} are more suited to predict properties of radial
excitations, since they are able to calculate directly their wave functions. However, as to {\it strong} decays, the elementary emission model does
not seem satisfactory (Roper $P_{11} \to N, \Delta \pi$ decays are wrong by a
factor {\it 10}). The $^3P_0$ {\it Quark pair creation model}  is better. It is good for $\Upsilon
' \to B\overline{B}$, $\psi ' \to D \overline{D}$. But, since it is
completely NR for both decay products, it cannot be quantitative for
$D'^{(*)} \to D^{(*)}\pi$.

Since decays of radial excitations are not either easily treated in {\it lattice}
QCD, this is an interesting challenge to theory.

\section{Confrontation of experiment, dynamical calculations and SV sum
rules : the example of $\rho^2$, $\tau_{1/2}$, $\mu_{\pi}^2$, $\rho_D^3$,
$\overline{\Lambda}$}
\label{sec4:leyaouanc}
An illustrative example of the above considerations is provided by the discussion
concerning an important set of quantities involved in heavy quark phenomenology :
$\rho^2$, $\tau_{1/2}$, $\mu_{\pi}^2$, $\rho_D^3$,
$\overline{\Lambda}$, defined at the end of the introduction. Our discussion
will overlap frequently with the ones given by N. Uraltsev and I. Bigi, yet our conclusions will be rather different or put a different emphasis on the various possibilities.

\subsection{Theoretical sources for these quantities}
\label{subsec4.2:leyaouanc}

\hspace*{\parindent} 1) {\bf Dynamical calculations}

$\rho^2$ and $\tau_{1/2}$ have
been calculated in {\it dynamical} approaches

- in QCDSR (``{\it three-point}'' sum rules)

${(\rho^2)} ^{\overline{MS}} \sim 0.84$ at 1.4 GeV,  with radiative corrections included 
to two loops ; $(\rho^2)^{SV}$(1 GeV) should not be far from this $\rho^2$ (superindex "SV" means a definition through the SV sum rules) .

$\tau_{1/2}(1) = 0.22$ in a first calculation, but later found to be enhanced by radiative corrections $\to 0.35$ ; however, a very large uncertainty 0.2 - 0.4 remains due to dependence on continuum threshold. Also, a much smaller result has been found with another interpolating current, and the slope of the form factor, $\tau_{1/2}(w)$, is found to be much flatter.

\noindent $ \begin{array}{l} \hbox{-in quark models \`a la Bakamjian-Thomas}\\
\hbox{with Godfrey-Isgur wave equation} \end{array} \left \{ 
\begin{array}{l} \rho^2 \approx 1 \\ \tau_{1/2}(1) = 0.225 \/
(\ll \tau_{3/2}(1) = 0.54)\end{array}  \right .$ 

The parameters $\overline{\Lambda}$ and $\mu_{\pi}^2(or~\lambda_1)$ can
also be calculated from QCDSR~:

- $\overline{\Lambda}$ from $m_b$, by a very safe QCDSR (2 points)
$\overline{\Lambda}^{SV}$(1 GeV) $\sim$ 0.7 GeV

- $\mu_{\pi}^2$ from 3-points QCDSR (perhaps overestimated) : $\mu_{\pi}^2 = 0.5$
GeV$^2$ + 0.17 GeV$^2$ $\sim$ 0.7 GeV$^2$

2) {\bf Bounds from SV sum rules} 

The "SV sum rules",
emphasized by Uraltsev,
relate basic heavy quark parameters of the ground state $\rho^2$, $\overline{\Lambda}$, $\mu_{\pi}^2$ , $\rho_D^3$ to sums over orbital excitations of terms of type :
$(\Delta E)_j^n (\tau_j)^2$ ($j=1/2,3/2$), cutoff at some excitation energy. One can derive lower bounds from these sum rules, with the bound given by the contribution of the lowest $j=1/2$ state.

$\rho^2$ and $\tau_{1/2}(1)$ are related to each other by the
bound :
$$(\rho^2)^{SV} \hbox{(1 GeV)} > {3 \over 4} + {\bf 3} \tau_{1/2}^2
(1)$$

\noindent the chosen cutoff 1 GeV being the estimated threshold of duality (it includes $\sim$ 2 or 3
levels). Analogous bounds relate $\tau_{1/2}$ to $\overline{\Lambda}$,
$\mu_{\pi}^2$, $\rho_D^3$ :
$$\begin{array}{l}  \overline{\Lambda}^{SV}\hbox{(1 GeV)} > 2 \Delta \left (
\displaystyle{{1 \over 2}} + {\bf 3} \tau_{1/2}^2(1) \right ) \\
\mu_{\pi}^2 \hbox{(1 GeV)}> \mu_G^2 + {\bf 9} \Delta^2
\tau_{1/2}^2(1) \\ 
(\rho_D^3)^{SV}\hbox{(1 GeV)} > \Delta \mu_{\pi}^2 \end{array} \begin{array}{l} \Delta \sim
0.35 \ \hbox{GeV} \hbox{: excitation energy}\\ \hbox{of $L=1$ $j=1/2$ states}
\\ \hbox{with respect to $L = 0$} \\ \hbox{at $m_Q = \infty$ ($\Delta$
conservatively small)}
\end{array}$$

NOTE THAT ALL THESE BOUNDS ARE VERY SENSITIVE TO $\tau_{1/2}(1)$, 
because of the large coefficients in front. Then, a safe determination of $\tau_{1/2}(1)$ by any means appears very crucial.

\subsection{Experimental sources}
\label{subsec4.3:leyaouanc}
\hspace*{\parindent}1) $(\rho^2)_{A_1}$ is not less than 1.2~; several
notably higher values (note the role of a {\it curvature}, bounded from
below, in enhancing $\rho^2$) although it seems a rather difficult
measure, with large errors. Now, $(\rho^2)^{SV}$(1 GeV) must not be very far from $\rho_{A_1}^2$
(taking into account various radiative  corrections), with $\pm 0.2$ of unknown
$1/m_Q$ corrections $\Rightarrow$ $(\rho^2)^{SV} > 1$.

2) $\tau_{1/2}(1)$. CLEO, and then BELLE with more
statistics have performed the measurement of $B^- \to D^{**}(0^+, 1_{1/2}^+$) $\pi^-$
with $D^{**}_{1/2}$ {\it well identified}. This fixes $\tau_{1/2}$ ($q^2 =
0$ or $w \sim 1.3$) assuming factorisation \footnote{Note added after the talk~: In this estimate
of $\tau_{1/2}$, we are also neglecting the color suppressed diagram present in this
class III decay. We thank BELLE collaboration (A. Bondar) for
underlining that.} {\bf (NB the unknown uncertainty on this assumption )} and small finite mass corrections . One finds $\framebox{$\tau_{1/2} (q^2 = 0) \sim 0.4$}$, consistently from $0^+$ and $1^+_{1/2}$ (0.3 from the lower bound); whence, with a reasonable assumption of a decreasing form factor $\tau_{1/2} (1) > 0.4$ ; $\tau_{3/2}$ is quite
consistent with $B \to D^{**}(1^+_{3/2})\ell \nu$ ($\sim 0.3$ at $q^2 =
0$)\footnote{For a more detailed discussion, see Appendix of our article hep-ph/0407176,v2}. A large $\tau_{1/2}$ is also indicated by the study of semileptonic decays at LEP (DELPHI). 

3) $\overline{\Lambda}$, $\mu_{\pi}^2$, $\rho_D^3$ can be obtained from fits to {\it moments}
of $b \to s \gamma$, $b \to c \ell \nu$

$$\begin{array}{l} \hbox{central}\\ \hbox{values}\\ \hbox{with large}\\
\hbox{errors}\end{array} \left \{ \begin{array}{l}
(\overline{\Lambda})^{SV}\hbox{(1 GeV) $\sim$ 0.6 to 0.72 GeV
(converted from $\overline{\Lambda}_{HQET}$)}\\ \mu_{\pi}^2 \sim 0.3 \
\hbox{to 0.45 GeV$^2$}\\ (\rho_D^3)^{SV} \sim 0.05 \end{array} \right . $$

\subsection{Consequences to be drawn from experiment}
\label{subsec4.4:leyaouanc}

1) Direct problems for dynamical calculations from exclusive decay measurements.

- The experimental $\rho_{A_1}^2$ seems much too large for QCDSR ; it is better, although
still large, for
Bakamjian-Thomas quark models which predict naturally a larger $\rho^2 \simeq 1$.

-The experimental result $\tau_{1/2}(q^2 = 0) \sim 0.4$ is too large for QCDSR  : the fastly decreasing form
factor leads to $0.2$ {\it at most} at $q^2 = 0$ ; it is also too large for the
Bakamjian-Thomas quark models : the more slowly decreasing form factor leads also to $0.2$ at $q^2
= 0$. \framebox{There is a discrepancy by a factor 2}\\
Note that the value at $w=1$ is not involved there.

\noindent 2) Problems generated by the bounds from SV sum rules on the moments. 

Since form factors are expected to be decreasing, we expect
that the $w=1$ value should be in fact larger than at $q^2=0$.
With a conservative $\tau_{1/2} (w=1) \sim 0.4$, we obtain already rather
strong lower bounds. The first bound above
gives :
\begin{equation}
(\rho^2)^{SV} > 1.25 
\end{equation}
which seems to confort the {\it large} fitted values of $\rho_{A_1}^2$, and once more disagrees with QCDSR estimate.

$\hbox{The other bounds give} \left \{ \begin{array}{l}
\overline{\Lambda}^{SV}\hbox{(1 GeV)} > 0.7\\ \mu_{\pi}^2\hbox{(1 GeV)}
> 0.53 \\ (\rho_D^3)^{SV} (\hbox{(1 GeV)} > 0.18 \end{array} \right .$
with $\mu_G^2$ taken from $B^*-B$~: 0.35 GeV$^2$ 

$\Rightarrow$ QCDSR estimates for $\overline{\Lambda}$ and $\mu_{\pi}^2$ are
in good agreement with these bounds.

$\Rightarrow$ Although {\it compatible} within large present
errors with these bounds, the {\it central} values of HQET parameters from fits to inclusive moments seem all {\it below} the corresponding lower bounds, which is somewhat strange ; in the case of $\mu_{\pi}^2$, there is a potential discrepancy. This discrepancy is not much reduced if we take the lower edge of error bars $\tau_{1/2} (q^2=0)=0.35$. {\it It is therefore crucial} to ascertain the determination
of $\tau_{1/2} (w=1)$ on the one hand, and the one of the moments on the other hand.

\section{Conclusions}
\label{sec5:leyaouanc}
\hspace*{\parindent} a) Heavy quark theory has
managed to establish a series of {\it very strong} relations between
apparently remote quantities

\noindent e.g.$\begin{array}{l} \hbox{\it strong lower bounds on} \end{array}$ $\left \{ \begin{array}{l}
\rho^2 \ \hbox {from} \ B \to D_{1/2}^{**}\pi, D^{**}-D~\mathrm{splitting}\\ \mu_{\pi}^2 \ \hbox {from} \ B \to D_{1/2}^{**}\pi, D^{**}-D, D^* -D 
 \end{array} \right .$  \\

b) Up to now, {\it analytical} dynamical methods (as opposed to the lattice approach
with Montecarlo calculations)
  are easy to handle, and give important conceptual insights, but they are not able to give very {\it safe} predictions, and it is difficult to foresee decisive progresses in this respect . At least, one wishes : - improvements of the {\it QCDSR} SVZ method by {\it explicit} inclusion of radial excitations, which seems important for heavy-light systems ; - formulations of {\it new quark
models} for {\it form factors}, combining the respective advantages of
Bakamjian-Thomas and Dirac equation. \\

c) In the present, we are faced with a rather large discrepancy with the class III $B \to D^{**}(j = 1/2)\pi$ experiment of both the main "analytical" methods , at least when neglecting the color suppressed contribution. \\

d) Some {\it lattice} calculations are then strongly needed 
to reduce the theoretical uncertainty on crucial
quantities like $\tau_{1/2}(1)$. A good precedent has been
$\widehat{g}$ \footnote{Since then, a first lattice calculation of $\tau_{1/2}(1)$ has been performed : D. Becirevic, B. Blossier et al., hep-lat/0406031. It yields a rather high value : 0.38(4)(statistical errors), but with rather unknown systematic errors}.

e) On the other hand, in view of the serious discrepancies observed above, checking the {\it experimental} determination of $B \to D^{**}(j = 1/2)\pi$ \footnote{Since then, BELLE has completed the measurement
of the corresponding class I decay, with neutral $B$ into charged $D^{**}$, which has the advantage of not being contaminated by the $D^{**}$ emission diagram. There is indeed a striking difference with the charged $B$ decay : 
the $j=1/2$ states seem strongly suppressed instead of being of roughly the same magnitude as the $j=3/2$.
Consequences from this result have to be drawn.} and
estimating the theoretical {\it finite mass corrections} is very important.
There is also the need to improve much the experimental determination of $\rho^2$,
the slope of Isgur-Wise function,
although this seems still more difficult, and the one of the moments in inclusive semileptonic decays .


\section*{Acknowledgments} I would like to thank warmly the organizers of the FPCP 
conference, who have allowed me to express for once personal views, which I have
usually to keep for discussions within our small group. I would like to thank in particular Jerome Charles and Luis Oliver, and G\'erard Bonneau. I would also like to apologize to the specialists of
QCD sum rules for some critical opinions which are not meant to diminish the
undisputable importance of this approach to hadronic physics.





\label{Le-YaouancEnd}
 
\end{document}